# EVENT CENTRIC MODELING APPROACH IN CO-LOCATION PATTERN ANALYSIS FROM SPATIAL DATA

Venkatesan.M[1], Arunkumar.Thangavelu[2], Prabhavathy.P[3]

[1 & 2]School of Computing Science & Engineering, VIT University, Vellore
{mvenkatesan@vit.ac.in,arunkumar.thangavelu@gmail.com}

[3]School of Information Technology & Engineering, VIT University, Vellore
{pprabhavathy@vit.ac.in}

## ABSTRACT

*Spatial co-location patterns are the subsets of Boolean spatial features whose instances are often located in close geographic proximity. Co-location rules can be identified by spatial statistics or data mining approaches. In data mining method, Association rule-based approaches can be used which are further divided into transaction-based approaches and distance-based approaches. Transaction-based approaches focus on defining transactions over space so that an Apriori algorithm can be used. The natural notion of transactions is absent in spatial data sets which are embedded in continuous geographic space. A new distance –based approach is developed to mine co-location patterns from spatial data by using the concept of proximity neighborhood. A new interest measure, a participation index, is used for spatial co-location patterns as it possesses an anti-monotone property. An algorithm to discover co-location patterns are designed which generates candidate locations and their table instances. Finally the co-location rules are generated to identify the patterns.*

## 1. INTRODUCTION

Huge amount of Geo-spatial data leads to definition of complex relationship, which creates challenges in today data mining research. Geo-spatial data can be represented in raster format and vector format. Raster data are represented in n-dimensional bit maps or pixel maps and vector data information can be represented as unions or overlays of basic geometric constructs, such as points, lines and polygons. Spatial data mining refers to the extraction of knowledge, spatial relationships, or other interesting patterns not explicitly stored in spatial data sets. As family of spatial data mining, spatial Co-location pattern detection aim to discover the objects whose spatial features/events that are frequently co-located in the same region. It may reveal important phenomena in a number of applications including location based services, geographic information systems, geo-marketing, remote sensing, image database exploration, medical imaging, navigation, traffic control and environmental studies. Some types of services may be requested in proximate geographic area, such as finding the agricultural land which is nearest to river bed. Location based service providers are very interested in finding what services are requested frequently together and located in spatial proximity. The co-location pattern and the rule discovery are part of spatial data mining process.

DOI: 10.5121/ijdms.2011.3311                       125

The differences between spatial data mining and classical data mining are mainly related to data input, statistical foundation, output patterns, and computational process. Co-location rules[12] are models to infer the presence of boolean spatial features in the neighborhood of instances of other boolean spatial features. Co-location rule discovery is a process to identify co-location patterns from large spatial datasets with a large number of boolean features. This paper discusses the detection of co-location pattern from the complex Geo-Spatial data by using event centric model approach. This paper is structured as follows: Section 2 discuses existing methods available to discover co-location pattern. Section 3 describes the model and the concepts of co-location pattern mining. Section 4 includes the proposed system design and co-location algorithm. Section 5 deals experimental execution of the co-location algorithm with the result implementation of each methodology. Section 6 summarizes the performance analysis and comparison our approach with the existing methods and Section 7 discusses the conclusions and future enhancements of the proposed system.

## 2. LITERATURE SURVEY

The Approaches to discovering co-location rules in the literature can be categorized into three classes, namely spatial statistics, data mining, and the event centric approach. Spatial statistics-based approaches use measures of spatial correlation to characterize the relationship between different types of spatial features using the cross-K function with Monte Carlo simulation and quadrant count analysis. Computing spatial correlation measures for all possible co-location patterns can be computationally expensive due to the exponential number of candidate subsets given a large collection of spatial boolean features. Data mining approaches can be further divided into a clustering-based map overlay approach and association rule-based approaches. Association rule-based approaches can be divided into transaction-based approaches and distance-based approaches Association rule-based approaches focus on the creation of transactions over space so that an apriori like algorithm [2] can be used. Transactions over space can use a reference-feature centric [3] approach or a data-partition approach [4]. The reference feature centric model is based on the choice of a reference spatial feature and is relevant to application domains focusing on a specific boolean spatial feature, e.g., incidence of cancer. Domain scientists are interested in finding the co-locations of other task relevant features to the reference feature [3]. Transactions are created around instances of one user specified reference spatial feature. The association rules are derived using the apriori[2] algorithm. The rules found are all related to the reference feature.

Defining transactions by a data-partition approach [4] defines transactions by dividing spatial datasets into disjoint partitions. A clustering-based map overlay approach [7], [6] treats every spatial attribute as a map layer and considers spatial clusters (regions) of point-data in each layer as candidates for mining associations. A distance-based approach [4],[5] was proposed called k-neighbouring class sets. In this the number of instances for each pattern is used as the prevalence measure [14], which does not possess an anti-monotone property by nature. The reference feature centric and data partitioning models materialize transactions and thus can use traditional support and confidence measures. Co-location pattern mining general approach [8] formalized the co-location problem and showed the similarities and differences between the co-location rules problem and the classic association rules problem as well as the difficulties in using traditional measures (e.g., support, confidence) created by implicit, overlapping and potentially infinite transactions in spatial data sets. It also proposed the notion of user-specified proximity neighborhoods[13][15] in place of transactions to specify groups of items and defined interest measures that are robust in the face of potentially infinite overlapping proximity neighborhoods.
A novel Joinless approach[9] for efficient collocation pattern mining uses an instance-lookup scheme instead of an expensive spatial or instance join operation for identifying collocation instances. A Partial join approach [9] is applied for spatial data which are clustered in neighbourhood area. Mining co-location patterns with rare spatial features [10] proposes a new measure called the maximal participation ratio (maxPR) and shown that a co-location pattern with a relatively high maxPR value corresponds to a collocation pattern containing rare spatial events. A novel order-clique-based approach [11] is used to mine maximal co-locations. In this paper distance based approach is used to find the co-location patterns from the spatial data. The participation index is used to prune the data to accept only the interesting patterns.

## 3. CO-LOCATION PATTERN MINING

Mining spatial co-location patterns is an important spatial data mining task. A spatial co-location pattern is a set of spatial features that are frequently located together in spatial proximity. Spatial co-location patterns represent relationships among events happening in different and possibly nearby grid cells. Co-location patterns are discovered by using any one of the model such as reference feature centric model, window centric model and event centric model. The prevalence measure and the conditional probability measure are called interesting measures used to determine useful co-location patterns from the spatial data. The interesting measures are defined differently in different models. Our approach is to find the co-location pattern from the spatial by using event centric model where the interesting measure is participation index

### 3.1 Event Centric Model Approach

The event centric model is relevant to applications like ecology where there are many types of Boolean spatial features. Ecologists are interested in finding subsets of spatial features likely to occur in a neighborhood around instance of given subsets of event types.

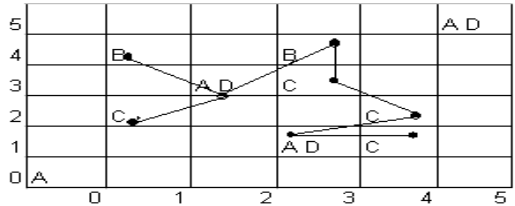

**Fig. 1.** Event Centric Model

Consider the figure 1, where the objective is to determine the probability of finding at least one instance of feature type B in the neighborhood of an instance of feature type A in figure 1. There are four instances of feature type A and two of them have some instances of type B in their 9 – neighbor adjacent neighborhoods. The conditional probability for the co-location rule is: Spatial feature A at location 1 → spatial feature type B in 9 neighbor neighborhood is 50%.Neighbourhood is an important concept in the event centric model.

### 3.2 Basic Concepts and Mathematical definition

For a spatial data set S, let F = { f1, . . . , fk} be a set of boolean spatial features. Let i = {i1, . . . , in} be a set of n instances in S, where each instance is a vector <instance-id, location, spatial features>. The spatial feature f of instance i is denoted by i. f .We assume that the spatial features of an instance are from F and the location is within the spatial framework of the spatial database. Furthermore, we assume that there exists a neighborhood relation R over pair wise instances in S.

**Case 1:** (A Spatial Data Set) Figure 2 shows a spatial data set with a spatial feature set F = {A, B, C, D}, which will be used as the running example in this paper. Objects with various shapes represent different spatial features, as shown in the legend. Each instance is uniquely identified by its instance-id. We have 18 instances in the database

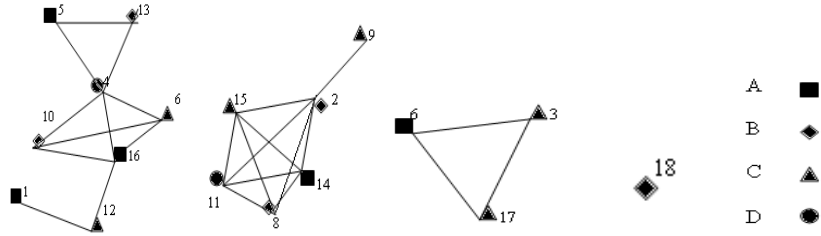

**Fig. 2.** Spatial data set

The objective of co-location pattern mining is to find frequently co-located subsets of spatial features. For example, a co-location {traffic jam, police, car accident} means that a traffic jam, police, and a car accident frequently occur in a nearby region. To capture the concept of "nearby," the concept of user-specified neighbor-sets was introduced. A neighbor-set L is a set of instances such that all pair wise locations in L are neighbors. A co-location pattern C is a set of spatial features, i.e., $C \subseteq F$. A neighbor-set L is said to be a row instance of co-location pattern C if every feature in C appears as a feature of an instance in L, and there exists no proper subset of L does so. We denote all row instances of a co-location pattern C by row set(C).

**Case 2:** (Neighbor-set, row instance and rowset) In Fig. 2, the neighborhood relation R is defined based on Euclidean distance. Two instances are neighbors if their Euclidean distance is less than a user specified threshold. Neighboring instances are connected by edges. For instance, {3, 6, 17}, {4, 5, 13}, and {4, 7, 10, 16} are all neighbor-sets because each set forms a clique. Here, we use the instance-id to refer to an object in Fig. 2. Additional neighbor-sets include {6, 17}, {3, 6}, {2, 15, 11, 14}, and {2, 15, 8, 11, 14}. {A, B,C, D} is a co-location pattern. The neighborhood-set {14, 2, 15, 11} is a row instance of the pattern {A, B,C, D} but the neighborhood-set {14, 2, 8, 15, 11} is not a row instance of co-location {A, B,C, D} because it has a proper subset {14, 2, 15, 11} which contains all the features in {A, B,C, D}.Finally, the rowset({A, B,C, D})= {{7, 10, 16, 4}, {14, 2, 15, 11},{14, 8, 15, 11}}.For a co-location rule R : A → B, the conditional probability cp(R) of R is defined as

$$\frac{|\{L \in \text{rowset}(A) \mid \exists L' \text{ s.t. } (L \subseteq L') \wedge (L' \in \text{rowset}(A \cup B))\}|}{|\text{rowset}(A)|} \qquad (1)$$

In words, the conditional probability is the probability that a neighbor-set in rowset(A) is part of a neighbor-set in rowset(A $\cup$ B). Intuitively, the conditional probability p indicates that, whenever we observe the occurrences of spatial features in A, the probability to find occurrence of B in a nearby region is p.

Given a spatial database S, to measure how a spatial feature f is co-located with other features in co-location pattern C, a participation ratio pr(C, f ) can be defined as

$$\text{pr}(C, f) = \frac{|\{r \mid (r \in S) \wedge (r.f = f) \wedge (r \text{ is in a row instance of } C)\}|}{\{r \mid (r \in S) \wedge (r.f = f)\}} \qquad (2)$$

In words, a feature f has a partition ratio pr(C, f ) in pattern C means wherever the feature f is observed, with probability pr(C, f ), all other features in C are also observed in a neighbor-set. A participation index was used to measure how all the spatial features in a co-location pattern are co-located. For a co-location pattern C, the participation index PI(C) = min $f \in C$ {pr(C, f )}.In words, wherever any feature in C is observed, with a probability of at least PI(C), all other features in C can be observed in a neighbor-set. A high participation index value indicates that the spatial features in a co-location pattern likely occur together. The participation index was used because in spatial application domain there are no natural "transactions" and thus "support" is not well-defined. Given a user-specified participation index threshold min_prev, a co-location pattern is called prevalent if $PI(C) \geq min\_prev$.

## 4. THE PROPOSED SYSTEM

### 4.1. The Proposed architecture

The proposed system input mainly consists of a satellite image which is processed to derive the co-ordinates item instances. The image is processed in Matlab where the instance is identified by colour identification. The co-ordinates are stored in a text file. The text file is processed to convert the co-ordinates into program readable format. The co-location algorithm is used to generate item sets from those co-ordinates. When the algorithm is applied the co-ordinates are mapped in a grid map. The distance between the instances is calculated. The 2-item sets are calculated by comparing the neighbouring grid spaces. The 2-item sets are pruned if patterns don't have minimum participation index. The non-pruned item sets are used to calculate the 3-item sets. The interesting patterns are identified after pruning depending on the participation index.

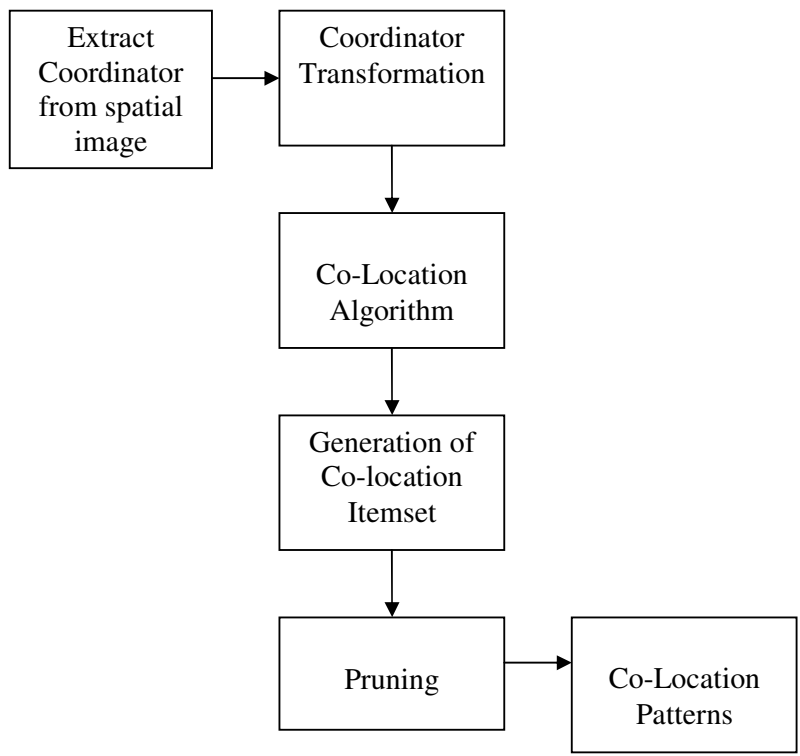

**Fig. 3.** General Architecture of Proposed System

### 4.2 Co-location Algorithm

One instance of an item is compared with all the instances of other item and checked for neighbourhood and participation index is found out and according to participation index the collocation pattern is predicted.

> 1. *Read the Coordinator values*
> 2. *for each element in first row of array*
>     *Compare each and every element in next row*
>     *Find different in grid coordinates*
>     *Check if difference leads to neighbor*
>     *Mark true in flag array if collocated*
>    *End loops*
> 4. *Calculate participation index by diving number of true s by total number of instances*
> 5. *Initialize pruning index*
> 6. *Compare participation index value and pruning index and consider only the item set that are abovethe pruning index.*
> 7. *The items that are pruned out in n-item set calculation are ruled out in n+1-item set calculation*
> 8. *End*

The collocation pattern is found out using participation index and using some pruning index value and some combinations are ruled out and final list of n-item set is proposed and only these combinations are taken to n+1-item set co-location analysis.

## 5. IMPLEMENTATION AND RESULT

The tools used for the implementation of co-location pattern mining are MATLAB 7.0, NET BEANS 6.1 and IE 7. The primary language of this experiment revolves around the largest open-source software JAVA. Hyper Text Markup Language (HTML) is used for displaying the result in a web browser.

### 5.1 Image Processing

In image processing we take an image which represent different objects and give it to MatLab for processing where each and every row is processed and different objects are identified and output is given to text file in raw format where x and y coordinates are separated by comma and each object is separated by type number

| X co-ord | Y co-ord | Type |
|---|---|---|
| 567 | 224 | 1 |
| 234 | 456 | 2 |
| 112 | 565 | 2 |
| 889 | 123 | 1 |
| 234 | 676 | 3 |
| 123 | 453 | 3 |

| 345 | 675 | 1 |

**Fig. 4.** Spatial Object Type in X and Y co-coordinator

### 5.2 Co-location pattern detection

Graphical User Interface allows user to browse the raw data file which is got through image processing. It has a method to take raw data as input and parse the whole text file and get x and y coordinates along with type value and write into another text file in a format that can be read by main program. This snapshot reflects the results of the co-location algorithm by grouping into item sets.

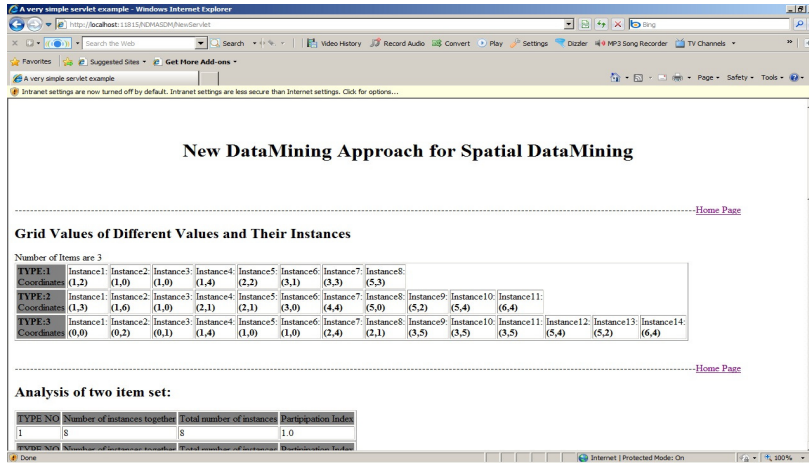

**Fig. 5.** Grid Coordinates

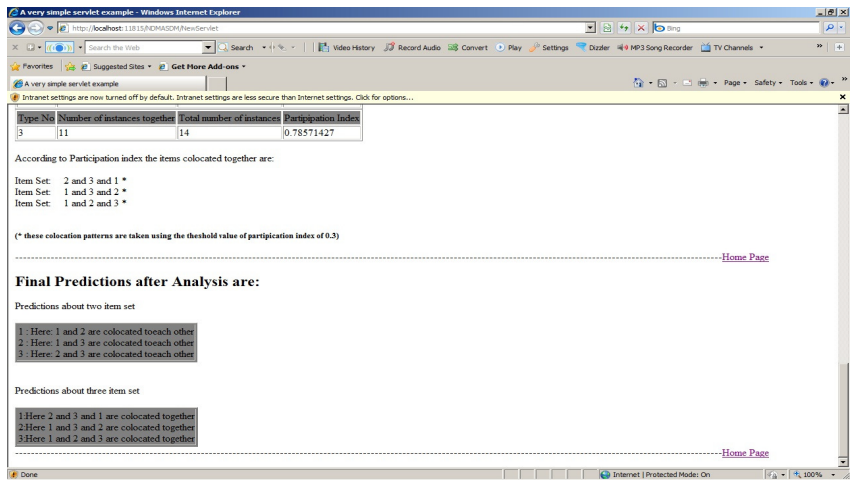

**Fig.6.** Co-located Pattern

This is the final predictions of the algorithm after processing the coordinates to various item sets. This represents the various entities which are collocated together.

## 6. PERFORMANCE ANALYSIS

Apriori algorithm generates huge number of candidate itemset to find frequent patterns and also scans the transaction database number of time to calculate the support count of the item. Also it

131

takes more time to generate more number of instances. Our Co-location algorithm takes minimum time to generate more number of instances in co-location pattern analysis. Figure 8.(a) and 8.(b) shows the comparison between apriori and collocation mining algorithm. Our approach does not need the constraint of "any point object must belong to only one instance" since we do not use the number of instances for a pattern as its prevalence measure. We propose the participation index as the prevalence measure, which possesses a desirable antimonotone property for effectively reducing the search space.

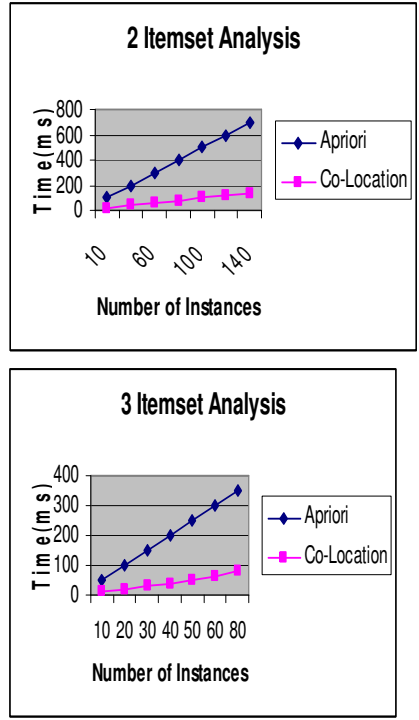

**Fig. 7.** Shows the comparison between Apriori and Co-location mining algorithm

Morimoto [4] provided an iterative algorithm for mining neighboring class sets with k + 1 feature from those with k features. In his algorithm, a nearest neighbor based spatial join was applied in each iteration. More specifically, a geometric technique, a Voronoi diagram, was used to take advantage of the restriction that "any point object must belong to only one instance of a k-neighboring class set." This algorithm considers a pure geometric join approach. In contrast, our co-location mining algorithm considers a combinatorial join approach in addition to a pure geometric join approach to generate size k+1 co-location patterns from size-k co-location patterns. Our experimental results show that a hybrid of geometric and combinatorial methods results in lower computation cost than either a pure geometric approach or pure combinatorial approach. In addition, we apply a multi resolution filter to exploit the spatial autocorrelation property of spatial data

## 7. CONCLUSIONS AND FUTURE ENHANCEMENTS

In this paper, we have discussed different approaches used to find the co-location pattern from the spatial data. We also have shown the similarities and differences between the co-location rules problem and the classic association rules problem. A new interest measure, a participation index, is used for spatial co-location patterns as it possesses an anti-monotone property. The new Co-location algorithm to mine collocation patterns from the spatial data was presented and analyzed. In future, the collocation mining problem should be investigated to account categorical and continuous data and also extended for spatial data types, such as line segments and polygons.